\documentclass[10pt,conference]{IEEEtran}
\IEEEoverridecommandlockouts

\usepackage{cite}
\usepackage{amsmath,amssymb,amsfonts}
\usepackage{algorithm}
\usepackage{algorithmic}
\usepackage{graphicx}
\usepackage{textcomp}
\usepackage{multirow}
\usepackage{xcolor}
\usepackage{subfig}
\usepackage{booktabs}
\usepackage[switch]{lineno}

\usepackage{pifont}
\usepackage{tcolorbox}
\usepackage[colorlinks=true, urlcolor=blue, linkcolor=blue, citecolor=blue]{hyperref}

\newcommand{\up}{\textcolor{green!60!black}{$\uparrow$}}
\newcommand{\down}{\textcolor{red!70!black}{$\downarrow$}}

\usepackage[normalem]{ulem}
\usepackage{color}

\usepackage{booktabs}
\def\BibTeX{{\rm B\kern-.05em{\sc i\kern-.025em b}\kern-.08em
    T\kern-.1667em\lower.7ex\hbox{E}\kern-.125emX}}

\newcommand{\method}{\textsc{DeepDiscovery}}

\begin{document}

\title{DeepDiscovery: A Location–Inference Framework for Task-Level Repository Understanding}

\author{
\IEEEauthorblockN{
Jiawei He\IEEEauthorrefmark{1},
Weisong Sun\IEEEauthorrefmark{2},
Mengyu Shi\IEEEauthorrefmark{3},
Jie Jia\IEEEauthorrefmark{1},
Tong Bian\IEEEauthorrefmark{4},
Xikai Yang\IEEEauthorrefmark{1}\thanks{Corresponding author: xikai.yxk@autonavi.com},
Dong Sun\IEEEauthorrefmark{1}
}
\IEEEauthorblockA{\IEEEauthorrefmark{1}AMAP, Alibaba Group}
\IEEEauthorblockA{\IEEEauthorrefmark{2}Nanyang Technological University}
\IEEEauthorblockA{\IEEEauthorrefmark{3}Nanjing University}
\IEEEauthorblockA{\IEEEauthorrefmark{4}University of Cambridge}
}



\maketitle

\begin{abstract}
Large language models have shown strong performance on software engineering (SE) tasks, yet understanding large industrial repositories remains challenging. 
Existing methods often retrieve only local fragments and fail to recover the broader task-relevant context needed for complex repository-level tasks.
We present {\method}, a  task-level repository-understanding framework for large language models (LLMs).
{\method} treats repository understanding as budgeted task-relevant context recovery: it first localizes high-confidence task anchors and then expands from them to recover broader implementation paths over a multi-relational repository structure.  
We evaluate {\method} in complementary settings: controlled method-level evaluation, organization-internal repository-understanding scenarios, and SWE-bench Verified for end-to-end task solving.
On 27 medium-scale tasks, {\method} achieves the best file recovery quality among five representative baselines without relying on persistent offline indexes or pre-built repository graphs.
On organization-internal industrial tasks from a production-scale integrated codebase ecosystem, including 27 medium-scale tasks and 40 large-scale tasks, {\method} improves Full Recall Rate across multiple AI coding systems, with absolute gains ranging from 1.6 to 9.2 percentage points on large subprojects and from 2.5 to 7.4 percentage points on medium-scale subprojects.
In a controlled end-to-end evaluation on SWE-bench Verified, replacing the baseline repository-understanding component with {\method} yields a 78.6\% Solve Rate, improving over the matched baseline by 8.2 percentage points. 
These results suggest that structured task-level context recovery is a reusable repository-understanding strategy for improving coding-agent performance on complex SE tasks.
\end{abstract}


\section{Introduction}
\label{sec:introduction}
Large language models (LLMs) have shown strong performance on a wide range of software engineering (SE) tasks, including code summarization~\cite{2025-LLM4CodeSum}, code generation~\cite{alharbi2026automatic}, program repair~\cite{zhang2024systematic}, 
repository-level question answering~\cite{yue2025survey}, and end-to-end task solving~\cite{zheng2025review}. 
However, strong performance on such tasks does not necessarily imply that current models can reliably understand large and evolving code repositories~\cite{alebachew2026beyond, pan2026archagent}. 
In realistic SE settings, successful task completion often requires reasoning not over a few isolated code fragments, but over a broader and more complete task-relevant context spanning interfaces~\cite{qiu2025locobench}, business logic~\cite{gao2026logicscan}, configuration~\cite{fu2024missconf}, tests~\cite{liu2024make}, and cross-module dependencies~\cite{he2026loogle}. 
Bridging this gap remains a fundamental challenge~\cite{du2025dependeval}.

The challenge is greater in large industrial repositories, which often contain many files, complex dependencies, mixed artifact types, and evolving organization~\cite{wang2026repomaster}. 
In such settings, existing methods often recover only limited local context. 
Semantic retrieval can return textually similar code fragments, but may miss structurally important yet lexically less salient entities~\cite{wen2026dependency}. 
Static dependency expansion can recover part of the explicit structure, but still struggles with many task-relevant links that are only implicitly expressed, such as configuration registration, dependency injection, event propagation, and cross-module constraints~\cite{luo2026gfm}. 
As a result, LLMs often see useful fragments without recovering the broader implementation context required by the task.

In practice, repository understanding also faces an important systems challenge beyond retrieval quality: repository freshness. 
Many repository-understanding methods rely on pre-built vector indexes~\cite{bevziuk2025vector}, static graphs~\cite{ma2025alibaba}, or other offline artifacts~\cite{li2025graphcodeagent}. 
While effective in relatively stable settings, such artifacts can be costly to maintain in industrial environments with frequent commits, branch switching, module evolution, and configuration changes. 
When synchronization lags behind repository evolution, retrieval quality and practical timeliness can both degrade. 
This makes repository understanding not only a question of what context is recovered, but also whether it can be recovered effectively under realistic freshness, budget, and deployment constraints.

To address these challenges, we propose {\method}, an executable Location–Inference framework for task-level repository understanding. 
Given a task and a current repository snapshot, {\method} first localizes high-confidence task anchors and then expands from them over explicit, implicit, and organizational relations to recover the implementation path needed by the task. 
To operate under context, latency, and freshness constraints, it constructs a metadata-first context package and promotes full file content only when local implementation details are needed. 
This design shifts repository understanding from one-shot fragment retrieval to budgeted recovery of structured task context.

We evaluate {\method} at the method, system, and end-to-end levels.
Across controlled repository-understanding benchmarks and a production-scale integrated codebase ecosystem, {\method} consistently improves task-relevant file recovery quality. 
In a controlled end-to-end evaluation on SWE-bench Verified, a system equipped with {\method} achieves a 78.6\% Solve Rate, outperforming the best baseline by 8.2 percentage points. 

The main contributions of this paper are as follows:
\begin{itemize}
    \item We formulate task-level repository understanding as budgeted structured-context recovery. The formulation emphasizes recovering the files, relations, and evidence needed to understand or complete a task, rather than retrieving locally similar fragments.

    \item We present {\method}, a Location–Inference framework that combines anchor localization, multi-relational expansion, and metadata-first context construction over the current repository snapshot. The framework is designed to operate without persistent offline indexes or pre-built repository graphs.

    \item We evaluate {\method} on controlled repository-understanding benchmarks and within multiple AI coding systems. The results show consistent improvements in Full Recall Rate (FRR) across medium-scale and large-scale industrial tasks.

    \item We assess downstream impact on SWE-bench Verified under a matched end-to-end setting. Replacing the baseline repository-understanding component with {\method} improves Solve Rate from 70.4\% to 78.6\%.
\end{itemize}

Desensitized code and experimental materials are available at \url{https://anonymous.4open.science/r/DeepDiscovery-4A4F}.

\section{Related Work}
\label{sec:related_works}

\subsection{Code Repository Understanding}
Recent work on LLM-based software engineering (SE) has moved beyond standalone code completion to tool-integrated and agentic systems that support repository exploration, iterative reasoning, and multi-step task execution~\cite{patel2025multi, chen2025acebench, yuan2025advancing, tian2025agentinit}. These systems have performed strongly on public benchmarks~\cite{jimenez2024swe, merrill2026terminal, he2026procbench}, suggesting promise for real-world SE tasks. At the same time, benchmark performance does not imply robust repository-scale understanding in realistic development settings, where tasks often involve cross-file dependencies, inter-module relationships, and long implementation paths~\cite{alebachew2026beyond, pan2026archagent, zhang-etal-2024-codeagent}. These limitations motivate methods that provide higher-quality repository-level context for downstream reasoning.

Retrieval-augmented generation (RAG) is a widely used paradigm for repository-level code understanding~\cite{yang2025empirical}. Typical methods partition repositories into file-level or chunk-level units, retrieve task-relevant content through semantic similarity, and provide the selected context to an LLM for downstream reasoning~\cite{du2024vul, guo2025empowering}. To improve on purely semantic retrieval, subsequent work incorporates static program structures such as call graphs, import graphs, and symbol reference graphs, yielding structure-augmented retrieval or GraphRAG-style pipelines~\cite{guo2025empowering, edge2024local, chinthareddy2026reliable}. These approaches generally improve cross-file coverage by exploiting explicit dependencies more effectively than pure vector retrieval~\cite{liu2026retrievalattention}.

However, most existing methods still frame repository understanding as locating relevant fragments. 
This is often insufficient for tasks that require broader implementation chains across files, modules, abstraction layers, configuration artifacts, and tests. 
The limitation is especially pronounced in industrial repositories, where many critical task-relevant links are only implicitly expressed.
Moreover, many retrieval-based pipelines rely on offline indexes or structural artifacts whose refresh cost can be high in fast-changing repositories~\cite{yang2025empirical, guo2025empowering, edge2024local, chinthareddy2026reliable}. 
These observations motivate methods that move beyond fragment retrieval to task-level context recovery under practical freshness and budget constraints.

Beyond standalone retrieval pipelines, recent research and industrial coding systems increasingly treat repository-scale SE as a system-level reasoning problem rather than pure context selection~\cite{wang2025burstgpt, he2026procbench}. In practice, coding assistants often combine repository search, multi-hop navigation~\cite{liu2025ds}, long-context reading~\cite{liusmooth}, execution tools~\cite{yuan2025easytool}, and iterative planning~\cite{lyu2024retrieve} to support complex tasks over large codebases. These systems show that end-to-end performance depends not only on retrieving relevant files, but also on connecting scattered evidence into a coherent task-specific reasoning process.

{\method} is closest to structure-augmented retrieval and graph-based repository exploration, but differs primarily in objective, output form, and execution constraints. Rather than optimizing only for query-relevant fragment ranking, it seeks to recover a compact implementation path containing task anchors, bridge entities, and supporting evidence under explicit budget and repository-freshness constraints. Repository structure is therefore used not only for ranking, but also for assembling a task-level context package that connects code, configuration, tests, and framework-mediated behavior.

\subsection{Code Feature Localization}
Code feature localization concerns identifying the code entities most likely to implement or constrain a target functionality or change request. In repository-scale SE tasks, localization is a critical first step because downstream reasoning depends on identifying reliable entry points before broader exploration begins. Existing retrieval-based methods typically localize candidates through lexical matching, embedding similarity, or query-conditioned ranking over files and code chunks~\cite{yang2025empirical, du2024vul}. Structure-augmented methods further improve localization by incorporating explicit dependency signals such as call, import, inheritance, and reference relations~\cite{guo2025empowering, edge2024local, chinthareddy2026reliable}.

However, accurate localization in large industrial repositories remains difficult when treated purely as a similarity-ranking problem. Many important entry points are only weakly expressed in local text and instead depend on naming conventions, framework idioms, configuration bindings, registration mechanisms, artifact roles, and directory organization. As a result, methods optimized for local relevance may still miss the high-confidence anchors needed for broader task resolution~\cite{alebachew2026beyond, pan2026archagent}. Our \emph{Location} stage is closely related to this line of work, but differs in performing environment-aware narrowing before candidate scoring and combining semantic evidence, structural summaries, rule-template matches, and task-conditioned artifact priors. This design aims to identify anchors that better support subsequent implementation-path recovery rather than merely ranking locally relevant files.

\section{Design of {\method}}
\label{sec:method}

\begin{figure*}[!t]
\centering
\includegraphics[width=\textwidth]{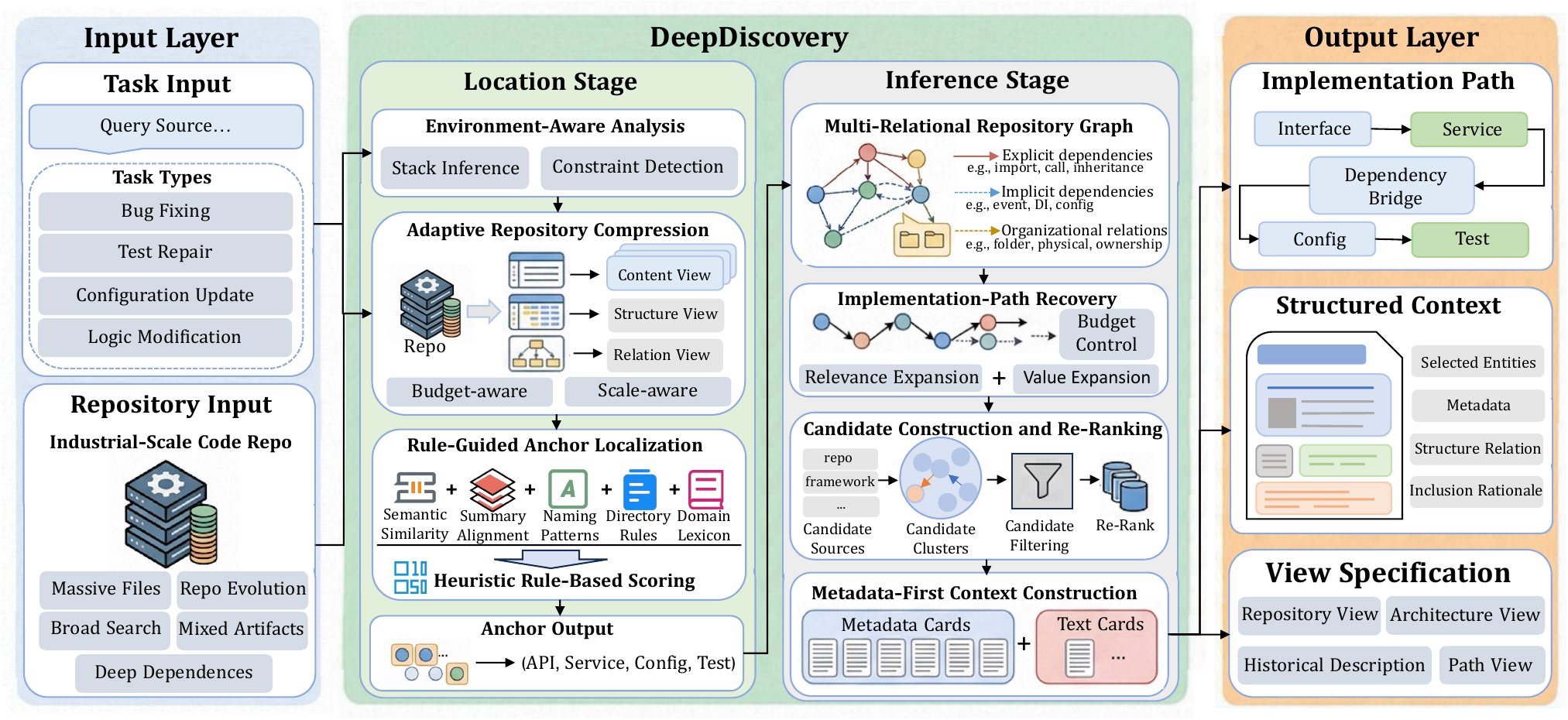}
\caption{Overview of {\method}. Given a task description and a large industrial repository, {\method} first localizes high-confidence task anchors and then expands from them over a multi-relational repository graph to recover a broader implementation path. The recovered evidence is organized into a metadata-first structured task-level context for downstream reasoning, rather than a flat list of retrieved files.
}
\label{fig:procbench_overview}
\end{figure*}

Figure~\ref{fig:procbench_overview} illustrates our {\method}. It takes as input a natural-language SE task and a current repository snapshot, and outputs a structured task context for downstream reasoning. The workflow has two stages: Location identifies high-confidence task anchors, and Inference expands from these anchors to recover the implementation path needed by the task.

We use the following terminology. 
A \textbf{repository entity} is a file-level artifact or finer-grained code, configuration, or test element selected during exploration. 
A \textbf{task anchor} is an entity with strong initial evidence of task relevance. 
An implementation path is the connected set of entities and relations needed to explain, constrain, modify, or validate task behavior. 
A \textbf{bridge entity} connects otherwise separate parts of this path, such as a configuration file, registration site, dependency-injection binding, or test artifact. 
In this work, implementation-path completeness is evaluated operationally through recovery of the expert-annotated task-relevant file set rather than through a separately annotated path structure.
A \textbf{structured context package} is the final output, comprising selected entities, their metadata or promoted text views, structural relations, and inclusion rationales.

\subsection{Problem Formulation}
Given a repository:
\begin{equation}
\small
\mathcal{R} = \{f_1, f_2, \dots, f_N\},
\end{equation}
where each $f_i$ is a repository artifact such as a source file, test file, or configuration file, and a natural-language task $q$, {\method} aims to construct a task-relevant structured context:
\begin{equation}
\small
\mathcal{B}(q)=\{(e_i, z_i, \rho_i)\}_{i=1}^{M},
\end{equation}
where $e_i$ is a selected repository entity, $z_i$ is its contextual representation, and $\rho_i$ records the structural rationale for including it. 
Depending on the stage of reasoning, $z_i$ may be either a metadata view or full-text content.

We formulate this process as \emph{budgeted context construction}. 
The goal is to recover a compact set of repository entities that jointly supports task understanding while respecting practical limits on context, tool usage, and latency.
Given a candidate entity space $\mathcal{V}$ induced from repository $\mathcal{R}$, we seek a context set with high task utility under a bounded exploration budget:
\begin{equation}
\label{eq3}
\small
\mathcal{C}^{\ast}(q)=
\arg\max_{\mathcal{C}\subseteq \mathcal{V}}
\operatorname{Util}(\mathcal{C}\mid q)
\quad
\text{s.t.}\quad
\operatorname{Cost}(\mathcal{C}) \le \Gamma_{\max},
\end{equation}
where  $\operatorname{Util}(\mathcal{C}\mid q)$ measures how well the constructed context supports task understanding, and $\operatorname{Cost}(\mathcal{C})$ captures the practical cost of exploration, context construction, and full-text loading.

This objective is not solved by a global optimizer. 
Instead, it serves as the design target for the executable Location--Inference workflow: candidate entities are selected when their expected utility justifies their budget cost. 
In our implementation, $\Gamma_{\max}$ is a task-time budget determined by the host system’s effective context window, tool-execution allowance, and latency target. 
Operationally, we instantiate $\Gamma_{\max}$ as a normalized budget over token usage and exploration steps, enabling comparable decision rules across repositories and tasks. 
This formulation therefore governs the scoring, expansion, and stopping decisions made during repository exploration.

\subsection{Repository Representation and Overall Framework}

{\method} represents the repository as a multi-relational graph
\begin{equation}
\small
\mathcal{G} = (\mathcal{V}, \mathcal{A}),
\end{equation}
where $\mathcal{V}$ denotes repository entities and $\mathcal{A}$ denotes relations among them. 
As shown in Figure~\ref{fig:procbench_overview}, the relation space combines three sources of structure:
(1) \emph{explicit dependencies}, such as imports, calls, inheritance, and references;
(2) \emph{implicit task-relevant links}, such as configuration-to-code mappings, registration sites, dependency-injection wiring, event or callback bindings, and test-to-implementation bridges; and
(3) \emph{organizational relations}, such as folder containment, module boundaries, ownership cues, and physical proximity.

Implicit links are extracted through a lightweight rule library that captures recurring repository idioms, such as configuration-to-code bindings, dependency-injection entries, event or callback mappings, and test-to-implementation bridges.
The rule inventory is fixed before evaluation and reused across tasks, so implicit-link construction is task-conditioned but not task-specifically authored.
Each rule is defined over observable repository evidence, including artifact roles, naming conventions, configuration keys, registration patterns, and directory proximity. 
The output of a rule is not treated as a deterministic dependency. 
Instead, it becomes a typed candidate relation with a confidence score, which is later combined with semantic and organizational evidence during expansion. 
Low-confidence links are expanded only when supported by additional evidence from other relation types or task cues.

In the \emph{Location} stage, the system narrows the search space and identifies a small set of high-confidence task anchors. 
In the \emph{Inference} stage, it expands from these anchors over the multi-relational repository graph, recovers a broader implementation path, and constructs a structured task-level context under the available budget.
Formally, let $\mathcal{H}\subseteq \mathcal{V}$ denote the task-anchor set. 
The overall process is implemented as:
\begin{equation}
\small
\hat{\mathcal{H}}=
\operatorname{Locate}(q,\mathcal{R},\Gamma_{\max}),
\qquad
\hat{\mathcal{C}}=
\operatorname{Infer}(\hat{\mathcal{H}}, q, \mathcal{G}, \Gamma_{\max}),
\end{equation}
In implementation, the total budget $\Gamma_{\max}$ is split into a localization budget $\Gamma_{\text{loc}}$ and a remaining inference budget $\Gamma_{\max}-\Gamma_{\text{loc}}$.
This decomposition reduces search cost while supporting repository-scale execution.

Algorithm~\ref{alg:deepdiscovery} summarizes the executable workflow in our implementation. 
To improve reproducibility and deployability, we make the main budget and stopping decisions explicit rather than leaving them implicit in LLM reasoning, while keeping several task-time scoring components in constrained executable forms instead of per-task tuning.
Thresholds are selected on a held-out development set and fixed across evaluation. Expansion also stops early if two consecutive actions introduce no new artifact role or bridge relation.

\begin{algorithm}[t]
    \caption{{\method}: Task-Level Repository Understanding}
    \label{alg:deepdiscovery}
    \small
    \begin{algorithmic}[1]
        \REQUIRE Task $q$, repository $\mathcal{R}$, budget $\Gamma_{\max}$
        \ENSURE Structured task context $\hat{\mathcal{C}}$
        \STATE \textbf{[Location]}\ Infer $\mathbf{c}_q$; build compressed view under $\Gamma_{\text{loc}}$; score via Eq.~(6); select $\hat{\mathcal{H}}$ (top-$K_a$)
        \STATE \textbf{[Inference]}\ $\mathcal{C}_0 \leftarrow \hat{\mathcal{H}}$,\ $\Gamma \leftarrow \Gamma_{\max} - \Gamma_{\text{loc}}$,\ $t \leftarrow 0$,\ $\text{stale} \leftarrow 0$
        \WHILE{$\Gamma > 0$ \textbf{and} $\text{stale} < 2$}
            \STATE Generate actions;\ $d^\ast \leftarrow \arg\max_d\ \operatorname{Priority}(d \mid \mathcal{C}_t, q)$ \hfill $\triangleright$ Eq.~(7)
            \STATE \textbf{if}\ $\operatorname{Priority}(d^\ast) < \tau$\ \textbf{then break}
            \STATE $\mathcal{C}_{t+1} \leftarrow \mathcal{C}_t \cup \operatorname{Expand}(d^\ast)$;\ update $\Gamma$, $t$, $\text{stale}$
        \ENDWHILE
        \STATE $z_e \leftarrow \mathbf{m}(e)$ or $\operatorname{FullText}(e)$ via Eq.~(8),\ for each $e \in \mathcal{C}_t$
        \RETURN $\hat{\mathcal{C}} \leftarrow \{(e,\, z_e,\, \rho_e)\}_{e \in \mathcal{C}_t}$
    \end{algorithmic}
\end{algorithm}

\vspace{-0.3em}
\subsection{Location: Task Anchor Localization}

The purpose of the Location stage is to identify a small set of high-confidence entry points for subsequent reasoning. 
As shown in Figure~\ref{fig:procbench_overview}, this stage consists of three steps: environment-aware analysis, adaptive repository compression, and rule-guided anchor localization.

\noindent\textbf{Environment-Aware Analysis.}
{\method} first infers task-relevant contextual factors such as the likely technology stack, module scope, artifact-role priors, and explicit task constraints. 
This step helps narrow the repository region that should be searched and suppresses obviously unrelated areas before deeper exploration begins.

\noindent\textbf{Adaptive Repository Compression.}
{\method} constructs a compressed repository view under the available budget. 
Rather than using a fixed summary format, {\method} adjusts the granularity of the repository view according to repository scale, structural complexity, and context budget. 
Under larger budgets, it preserves richer module and interface cues; under tighter budgets, it prioritizes compact structural summaries and representative entities.

\noindent\textbf{Rule-guided Anchor Localization.}
Finally, the system ranks candidate entities using four executable signals:
\begin{equation}
    \small
    \begin{aligned}
    \operatorname{Score}_{\text{loc}}(v\mid q)
    ={}& \alpha \tilde{s}_{\text{sem}}(q,v)
    +\beta \tilde{s}_{\text{sum}}(q,\mathcal{S}_v) \\
    &+\gamma \tilde{s}_{\text{rule}}(v,q,\mathbf{c}_q)
    +\delta \pi(\operatorname{type}(v)\mid q).
    \end{aligned}
\end{equation}
where $\tilde{s}_{\text{sem}}(q,v)$ combines embedding similarity and BM25-style lexical matching between the task and the metadata view of entity v; $\tilde{s}_{\text{sum}}(q,\mathcal{S}_v)$ measures alignment with the compressed structural view of v using a constrained LLM judgment over fixed fields, such as task relevance, structural role, and evidence sufficiency; $\tilde{s}_{\text{rule}}(v,q,\mathbf{c}_q)$ aggregates matched rule templates such as naming and directory patterns, framework-specific keywords, and domain-lexicon cues; and $\pi(\operatorname{type}(v)\mid q)$ is a task-conditioned prior over artifact roles. The LLM-assisted judgment is used only as a bounded scoring component over preconstructed metadata, not as an unconstrained repository exploration step. Its prompt uses fixed output fields and contributes to the execution-token cost reported in Table~\ref{tab:rq1_baselines}.
In implementation, matched rule templates are injected into a constrained scoring prompt with fixed output fields. 
The coefficients $\alpha, \beta, \gamma, \delta$ are selected from a small set of predefined configurations based on repository scale and localization ambiguity, and are fixed before evaluation. 
No coefficient, threshold, or rule template is tuned on individual evaluation tasks.
This scoring policy is used as an executable task-time heuristic rather than a learned optimizer.
The prior $\pi(\operatorname{type}(v)\mid q)$ uses language-specific artifact-role characteristics, keeping anchor scoring executable while allowing limited task-time adaptation.
The top-ranked entities are selected as task anchors for the Inference stage.

\subsection{Inference: Multi-Relational Implementation-Path Recovery}

Given the task anchors produced by Location, the Inference stage incrementally recovers a broader task-relevant implementation path. 
As shown in Figure~\ref{fig:procbench_overview}, this stage operates over a multi-relational repository graph and consists of implementation-path recovery, candidate construction, re-ranking, and structured context construction.

At each step, {\method} considers candidate expansion actions from the current context. 
An action may follow explicit dependencies, implicit task-relevant links, organizational proximity, or local-search results. 
Rather than expanding all reachable neighbors uniformly, the system prioritizes expansion actions using a budget-aware score:
\begin{equation}
\small
\operatorname{Priority}(d\mid \mathcal{C}_t,q)
=
\frac{\operatorname{Gain}(d\mid \mathcal{C}_t,q)}{\operatorname{Cost}(d)}.
\end{equation}
Here, $\operatorname{Gain}(d\mid \mathcal{C}_t,q)$ estimates how much action $d$ improves task relevance and implementation-path completeness, while $\operatorname{Cost}(d)$ estimates the additional exploration and context-construction cost introduced by that action under the same normalized budget accounting used for $\Gamma_{\max}$.
In implementation, each action $d$ is defined at relation-edge granularity.

$\operatorname{Gain}(d\mid \mathcal{C}_t,q)$ combines the relevance of newly reachable entities, the likelihood that the action bridges currently separated but jointly necessary parts of the implementation path, and the extent to which it helps recover still-missing artifact roles such as configuration, registration, tests, or interface-to-implementation links in an executable heuristic form.
$\operatorname{Cost}(d)$ reflects the additional context and execution overhead introduced by the action, including estimated token usage and exploration latency.
Expansion stops when the remaining budget is exhausted, when the best available priority falls below a threshold, or when no candidate action is predicted to add sufficient task-relevant coverage.
The selected actions are used to construct candidate clusters of related entities around the current anchors and partially recovered path, which are then filtered and re-ranked before being added to the task context.

\subsection{Metadata-First Context Construction}

The final part of the Inference stage constructs task context in a metadata-first manner, as illustrated by the metadata-card and text-card views in Figure~\ref{fig:procbench_overview}. 
The key idea is to preserve structural coverage while loading full text only when necessary. 
This follows a coverage-before-detail principle: for repository-level tasks, exposing the full text of a few locally relevant files too early can exhaust the context budget before the system identifies bridge entities, configuration artifacts, or tests. 
Metadata cards retain path, role, neighboring relations, and inclusion rationale at much lower cost, allowing DEEPDISCOVERY to maintain a broader structural view. 
Full text is promoted only for entities whose local implementation details are likely to matter for downstream reasoning or editing.

For each selected entity $e$, {\method} chooses between a metadata view $\mathbf{m}(e)$ and full-text content $\operatorname{FullText}(e)$:
\begin{equation}
    \small
    z_e=
    \begin{cases}
    \mathbf{m}(e), & \text{if } u(e\mid q,\mathcal{C}_t) < \kappa,\\
    \operatorname{FullText}(e), & \text{otherwise},
    \end{cases}
\end{equation}
where $u(e\mid q,\mathcal{C}_t)$  is a task-time promotion score indicating whether full-text inspection is necessary. 

In implementation, $u(e\mid q,\mathcal{C}_t)$ is computed from lightweight symbolic features, including artifact role, relation type, anchor distance, matched task cues, and whether the entity bridges currently disconnected parts of the recovered path. 
A constrained LLM judgment is used only to choose among fixed promotion labels, such as metadata-only, partial-text, or full-text, under a fixed-form decision policy defined before evaluation.
This fixed-form policy avoids a learned controller and keeps promotion decisions auditable; its token usage is included in the execution cost. 
After promotion, {\method} compresses and serializes the selected evidence into a structured context package for the downstream model. 
The package contains selected entities, metadata summaries, promoted text spans, structural relations, and inclusion rationales. 
Each metadata card records the repository path, artifact type, inferred role, local summary, matched task cues, neighboring relations, bridge entities, and reason for inclusion.

\section{Evaluation}
\label{sec:evaluation}
We evaluate {\method} from three perspectives: controlled method-level comparison, system-level repository-understanding evaluation in real AI coding systems, and end-to-end impact on SWE-bench Verified. We further include ablation, efficiency, and error analysis.

\subsection{Research Questions}

\textbf{RQ1:} 
How does {\method} compare with representative repository-understanding baselines in recovery quality and practical cost on medium-scale benchmarks?

\textbf{RQ2:} 
How does {\method} affect repository-understanding quality in real AI coding systems, especially in terms of fully recovering task-relevant files?

\textbf{RQ3:} 
How do the major components of {\method} contribute to task-level context recovery?

\textbf{RQ4:} 
How does {\method} affect end-to-end task solving on realistic software engineering tasks?

\textbf{RQ5:} 
How robust are the gains of {\method} across AI coding systems, project scales, and model backends?

\subsection{Experimental Setup}
\label{sec:experimental_setup}

\noindent\textbf{a) Benchmarks and evaluation scenarios.}
We adopt complementary evaluation settings. 
This choice is necessary because no existing public benchmark directly captures the primary objective of this work: task-level repository understanding. 
Specifically, this objective requires recovering the complete set of task-relevant files needed for repository-scale reasoning.
Public benchmarks such as SWE-bench Verified are valuable for end-to-end evaluation and for assessing downstream task-solving performance. 
However, they generally do not provide expert-annotated gold sets of relevant files. 
Such annotations are necessary for directly evaluating repository understanding.
In contrast, organization-internal repositories allow us to examine whether the same general framework remains effective in more integrated environments. 
These environments are characterized by multi-subproject structures, heterogeneous artifacts, implicit framework-mediated relations, and cross-module dependencies.
Therefore, our benchmark is designed to complement existing public end-to-end benchmarks. It is not intended to replace them.

The medium-scale and large benchmarks are drawn from a production-scale integrated codebase ecosystem with multiple large subprojects and heterogeneous business modules. The ecosystem contains 2.67M lines of code and 25K+ files across structurally distinct subprojects with different scopes, dependency patterns, and organizational boundaries. The medium-scale benchmark supports rapid validation on smaller subprojects, while the large benchmark covers larger and more complex subprojects and business scenarios from the same ecosystem. For RQ1, we use 27 medium-scale tasks with 135 manually annotated gold relevant files. For repository-understanding evaluation, we use these 27 medium-scale tasks together with 40 large-subproject tasks under the same annotation protocol. For RQ4, we use SWE-bench Verified under a controlled comparison protocol.

\textit{Gold annotations.} 
The gold relevant-file sets for repository-understanding evaluation were prepared independently by domain experts prior to the development of {\method} and were not created for this study. 
For each task, annotators identified the minimal file set required for understanding and completing the task, including implementation, bridge, configuration or registration, and test files where necessary. 
Ambiguous cases were resolved through joint review based on whether the file was necessary to explain, constrain, or validate the intended change. 
The annotations were used only for evaluation and were not exposed to the systems under assessment.
These annotations serve as a practical evaluation target for task-level repository understanding and are used exclusively for evaluation purposes. 
They are never exposed to the systems under assessment. Because the annotation process predates {\method}, the benchmark is method-independent and less prone to post hoc evaluation bias, although slight expert variation remains possible. 
We release the {\method} implementation, rule specifications, prompts, configuration files, evaluation scripts, and public-benchmark protocols to support independent reproduction and inspection. 
For industrial evaluations involving organization-internal repositories, we provide desensitized task descriptions, annotation guidelines, aggregate repository statistics, gold-file metadata, and per-run outputs where permitted. 
This separation allows readers to reproduce the public experiments directly and audit the industrial protocol without requiring access to proprietary source code.

\noindent\textbf{b) Compared systems and baselines.}
To evaluate the applicability of {\method} in realistic development settings, we consider six representative AI coding systems: Cline, Cursor, Claude Code, Codex, OpenCode, and Qoder. 
Cline and Cursor use \texttt{Claude Sonnet 4.5}, Claude Code uses \texttt{Claude Opus 4.6}, Codex and OpenCode use \texttt{‌GPT-5.4}, and Qoder uses \texttt{‌Qwen3.5-Plus}. 
These systems cover diverse workflows, including search-driven and tool-orchestration-based systems, as well as systems with stronger underlying models.

At the method level, we compare {\method} with five representative repository-understanding baselines: DeepWiki, CodeWiki, RAG, GraphRAG, and AST+GraphRAG~\cite{chinthareddy2026reliable}. 
To reduce confounding from host-system or model differences, all methods in RQ1 are run on Claude Code with the same foundation model, \texttt{Claude Opus 4.6}. 
Each configuration is executed three times.

\textit{Controlled comparison and fairness.}
For the RQ1 method-level comparison, the proposed method and five baselines are evaluated under the same task set, execution environment, host system, foundation model, context limit, and repetition protocol.
We separate preprocessing from task-time execution cost. For RAG, GraphRAG, and AST+GraphRAG, preprocessing includes index construction, graph construction, AST parsing, and repository refresh. 
For DeepWiki, CodeWiki, and {\method}, it includes summary construction, compression, and other preparatory operations under the same accounting rule. 
In large and evolving repositories, preprocessing latency, refresh overhead, and token cost are part of practical method quality and are therefore included in the evaluation scope. 
The results should be interpreted as a controlled comparison of reusable repository-understanding pipelines under a shared model, task set, context limit, and execution policy, rather than as a universal ranking over all possible retrieval, indexing, graph-construction, or reranking variants.
For the end-to-end evaluation on SWE-bench Verified, the baseline system and the {\method}-enhanced system use the same model, prompting strategy, tool interface, execution policy, and evaluation environment; only the repository-understanding component, including context construction, is changed.

\textit{Practical scope of comparison.}
The compared systems are heterogeneous practical coding agents rather than research backbones sharing a common architecture. We therefore study {\method} both as a standalone method under a unified host system in RQ1 and as an enhancement module inside realistic coding agents in the system-level evaluation.

\begin{table*}[t]
\caption{Method-level comparison on the 27 medium-scale benchmarks. All methods are run on top of Claude Code with \texttt{Claude Opus 4.6} and executed three times. FRR is reported as mean (min--max) across runs; MR and MP are reported as ranges across runs. Preprocessing and execution metrics refer only to repository-understanding cost rather than full-task cost.}
\label{tab:rq1_baselines}
\centering
\footnotesize
\setlength{\tabcolsep}{4pt}
\resizebox{\textwidth}{!}{%
\begin{tabular}{lccccccc}
\toprule
Method & FRR & MR & MP & Avg. Preprocess Time (h) & Avg. Execution Time (s) & Avg. Preprocess Tokens (Billion) & Avg. Execution Tokens \\
\midrule
Claude Code(Native)             & 88.9\% (85.2\%--92.6\%) & 90.4\%--95.6\% & 10.2\%--10.9\% & 0   & 0 & 0  & 0 \\
DeepWiki             & 88.0\% (83.4\%--91.8\%) & 89.6\%--94.8\% & 10.8\%--11.3\% & 0.0   & 7.2 & 0  & 4{,}927 \\
CodeWiki             & 88.2\% (84.0\%--90.2\%) & 90.4\%--95.6\% & 10.6\%--11.1\% & 0.0   & 8.0 & 0  & 10{,}730 \\
RAG                  & 84.4\% (81.5\%--88.9\%) & 85.8\%--91.8\% & 10.5\%--10.9\% & $\approx 7.0$  & 57.2 & 2.26 & 24{,}991 \\
GraphRAG             & 90.2\% (86.4\%--92.6\%) & 88.6\%--96.2\% & 10.1\%--10.6\% & $\approx 9.0$  & 63.8 & 3.14 & 28{,}338 \\
AST+GraphRAG         & 90.4\% (87.2\%--94.4\%) & 90.2\%--94.6\% & 10.0\%--10.5\% & $\approx 9.0$ & 44.8 & 3.98 & 29{,}677 \\
{\method}            & 92.6\% (88.9\%--96.3\%) & 90.4\%--96.3\% & 10.8\%--11.1\% & 0.0   & 13.2 & 0 & 8{,}826 \\
\bottomrule
\end{tabular}%
}
\end{table*}

\subsection{Metrics}
\label{sec:eval_protocol}

\paragraph{Repository-understanding metrics.}
For repository-understanding quality, we use \textbf{Full Recall Rate (FRR)} as the primary metric. 
Let $\mathcal{Q}$ denote the task set, and let $\mathcal{Y}(q)$ and $\hat{\mathcal{Y}}(q)$ denote the gold and predicted relevant-file sets for task $q$, respectively. 
A task $q \in \mathcal{Q}$ is considered fully recalled if:
\begin{equation}
\small
\hat{\mathcal{Y}}(q)\supseteq \mathcal{Y}(q).
\end{equation}
The Full Recall Rate is defined as:
\begin{equation}
\small
\operatorname{FRR}=
\frac{1}{|\mathcal{Q}|}
\sum_{q\in \mathcal{Q}}
\mathbf{1}\!\left[\hat{\mathcal{Y}}(q)\supseteq \mathcal{Y}(q)\right].
\end{equation}

We use FRR as the primary repository-understanding metric because the target problem is completeness-oriented. 
For many repository-level tasks, missing a bridge file, configuration artifact, registration site, or test file can make the recovered context insufficient, even when several locally relevant implementation files are retrieved. 
FRR therefore captures whether a method recovers the full gold file set for a task. 
At the same time, FRR is intentionally strict: missing a single necessary bridge or validation file counts as failure for that task. 
We therefore interpret FRR together with Micro Recall, Micro Precision, execution cost, and downstream end-to-end results rather than as a standalone measure of practical utility. 
MR measures aggregate file-level coverage across tasks, while MP reflects how much non-gold context is introduced.
We interpret these metrics together with execution cost, since a useful repository-understanding method should improve task-level completeness without relying on excessive context expansion. 
Unless otherwise stated, all methods are evaluated under the same host-system context limit and task-time execution policy.
In our controlled comparisons, all methods operate under the same host-system context limit and task-time execution policy, so FRR gains cannot arise from unbounded file accumulation. Let $H$ denote the total number of correctly recovered files across all tasks, $F$ the total number of gold relevant files, and $P$ the total number of predicted files.

\paragraph{End-to-end metric.}
For the paired SWE-bench comparison, we additionally apply the McNemar test on per-instance solved/unsolved outcomes between the baseline and the {\method}-enhanced system.
For this end-to-end software engineering performance, let $Q$ denote the total number of tasks and $Q_{\text{solved}}$ the number of successfully solved tasks. 
We define \textbf{Solve Rate (SR)} as:
\begin{equation}
\small
\operatorname{SR}=
\frac{Q_{\text{solved}}}{Q}.
\end{equation}

\paragraph{Cost metrics.}
To assess practical efficiency, we additionally report cost-related metrics for the method-level comparison: \textbf{Avg. Preprocess Time}, \textbf{Avg. Execution Time}, \textbf{Avg. Preprocess Tokens}, and \textbf{Avg. Execution Tokens}. 
Here, preprocessing refers to method-specific preparation steps before the task-time repository-understanding stage, such as repository summary construction, vector-index construction, graph construction, AST parsing, or related structural preprocessing. 
Execution refers only to the repository-understanding step itself, including task-time localization, retrieval, expansion, reranking, and context construction for the current task, rather than the total time or total token usage for completing the full software engineering task.

\paragraph{Metric derivation and reporting.}
All repository-understanding metrics are computed from run-level count statistics. For the large-subproject benchmark, we use $T=40$ tasks and $F=240$ gold relevant files; for the medium-scale benchmark, $T=27$ and $F=135$. Each configuration is executed three times. FRR is reported as mean (min--max) across runs, while MR and MP are reported as min--max across runs.
Unless otherwise stated, repository-understanding metrics are computed against expert-annotated gold relevant-file sets, while end-to-end Solve Rate is computed under the benchmark's official execution and verification protocol.

\subsection{Experimental Results}
\noindent\textbf{(1) RQ1: Method Comparison on Medium-Scale Tasks.}

To answer this question, we compare {\method} with five representative repository-understanding baselines: DeepWiki, CodeWiki, RAG, GraphRAG, and AST+GraphRAG, on the 27 medium-scale benchmarks under a unified controlled setting.
Table~\ref{tab:rq1_baselines} summarizes the results. 
Overall, {\method} achieves the highest observed FRR (92.6\%) among all compared methods while requiring no persistent offline index or graph construction.
From a quality perspective, {\method} outperforms all five baselines in FRR, exceeding AST+GraphRAG (90.4\%), GraphRAG (90.2\%), CodeWiki (88.2\%), DeepWiki (88.0\%), and RAG (84.4\%). 
Graph-based baselines improve over pure RAG, but still remain below {\method}; RAG performs worst, especially in MR.
{\method} is also competitive in MR and MP, indicating that its higher FRR does not come with a disproportionate precision penalty.

\begin{table*}[t]
\caption{Repository-understanding results of AI coding systems on large and medium-scale subprojects. Large-subproject results use $T{=}40$ tasks and $F{=}240$ gold relevant files; medium-scale results use $T{=}27$ tasks and $F{=}135$ gold relevant files. Full Recall is reported as mean (min--max) over three runs; Micro Recall and Micro Precision as min--max. Arrows show changes from the native setting, and +DD denotes the {\method}-enhanced setting.}
\label{tab:main_compare}
\centering
\footnotesize
\setlength{\tabcolsep}{3pt}
\resizebox{\textwidth}{!}{%
\begin{tabular}{ll|ccc|ccc}
\toprule
\multirow{2}{*}{System} & \multirow{2}{*}{Setting}
& \multicolumn{3}{c|}{Large}
& \multicolumn{3}{c}{Medium} \\
\cmidrule(lr){3-5}\cmidrule(lr){6-8}
&
& Full Rec.
& Mic. Rec.
& Mic. Prec.
& Full Rec.
& Mic. Rec.
& Mic. Prec. \\
\midrule
Cline & Native & 72.5\% (67.5\%--77.5\%) & 75.0\%--82.5\% & 8.9\%--9.5\% & 79.0\% (74.1\%--85.2\%) & 82.2\%--90.4\% & 9.6\%--11.2\% \\
Cline & + DD & \textbf{81.7\% (77.5\%--85.0\%) \up} & \textbf{82.1\%--88.3\% \up} & \textbf{9.4\%--10.2\% \up} & \textbf{86.4\% (81.5\%--92.6\%) \up} & \textbf{89.6\%--96.3\% \up} & \textbf{10.3\%--11.9\% \up} \\
\midrule

Cursor & Native & 82.5\% (77.5\%--87.5\%) & 83.8\%--89.2\% & 8.9\%--9.5\% & 85.2\% (81.5\%--88.9\%) & 87.4\%--93.3\% & 10.0\%--10.5\% \\
Cursor & + DD & \textbf{85.0\% (80.0\%--90.0\%) \up} & \textbf{87.5\%--92.1\% \up} & \textbf{9.1\%--9.6\% \up} & \textbf{88.9\% (85.2\%--92.6\%) \up} & \textbf{91.9\%--96.3\% \up} & \textbf{10.2\%--10.4\% \down} \\
\midrule

Claude Code & Native & 85.0\% (80.0\%--90.0\%) & 86.7\%--92.1\% & 9.2\%--9.7\% & 88.2\% (86.4\%--92.2\%) & 90.4\%--95.6\% & 10.2\%--10.9\% \\
Claude Code & + DD & \textbf{87.5\% (82.5\%--92.5\%) \up} & \textbf{88.3\%--93.3\% \up} & \textbf{9.1\%--9.3\% \down} & \textbf{91.4\% (86.6\%--94.0\%) \up} & \textbf{90.4\%--96.3\% \up} & \textbf{10.8\%--11.1\% \up} \\
\midrule

Codex & Native & 84.2\% (77.5\%--90.0\%) & 85.8\%--91.2\% & 9.0\%--9.6\% & 88.9\% (85.2\%--92.6\%) & 89.6\%--94.8\% & 10.0\%--10.8\% \\
Codex & + DD & \textbf{85.8\% (80.0\%--90.0\%) \up} & \textbf{85.4\%--90.8\% \down} & \textbf{9.1\%--9.6\% \up} & \textbf{91.4\% (86.6\%--94.0\%) \up} & \textbf{91.9\%--96.3\% \up} & \textbf{10.2\%--11.0\% \up} \\
\midrule

OpenCode & Native & 82.5\% (77.5\%--87.5\%) & 85.0\%--90.4\% & 9.0\%--9.5\% & 87.7\% (81.5\%--92.6\%) & 88.9\%--94.8\% & 10.0\%--10.8\% \\
OpenCode & + DD & \textbf{87.5\% (82.5\%--92.5\%) \up} & \textbf{89.6\%--94.6\% \up} & \textbf{9.1\%--9.4\% \down} & \textbf{90.2\% (82.6\%--91.2\%) \up} & \textbf{94.1\%--97.8\% \up} & \textbf{10.5\%--11.0\% \up} \\
\midrule

Qoder & Native & 82.5\% (77.5\%--87.5\%) & 84.2\%--89.6\% & 8.9\%--9.4\% & 85.2\% (81.5\%--88.9\%) & 88.1\%--94.1\% & 9.9\%--10.7\% \\
Qoder & + DD & \textbf{84.2\% (80.0\%--87.5\%) \up} & \textbf{86.7\%--91.2\% \up} & \textbf{9.0\%--9.5\% \up} & \textbf{88.9\% (85.2\%--92.6\%) \up} & \textbf{91.1\%--95.6\% \up} & \textbf{10.0\%--10.7\% \up} \\
\bottomrule
\end{tabular}%
}
\end{table*}

The efficiency results further support this conclusion. {\method}, DeepWiki, and CodeWiki require no offline preprocessing, whereas RAG, GraphRAG, and AST+GraphRAG incur substantial preparation cost. 
RAG requires about 7.0 hours of preprocessing and 2.26 billion preprocessing tokens on average, while GraphRAG and AST+GraphRAG each require about 9.0 hours, with 3.14 and 3.98 billion preprocessing tokens, respectively. 
At task time, DeepWiki and CodeWiki are the fastest methods, but they underperform {\method} in FRR. More importantly, all three RAG-style pipelines are markedly slower than {\method}, requiring 57.2, 63.8, and 44.8 seconds, respectively, versus 13.2 seconds for {\method}. 
Although {\method} uses more execution tokens than DeepWiki and CodeWiki, it still uses fewer than AST+GraphRAG while achieving the best FRR. 
In summary, {\method} achieves the highest FRR under the reported controlled setting while avoiding offline preprocessing. 
This result should not be interpreted as a standalone retrieval-quality improvement, but as a stronger recall–cost trade-off for task-level context recovery under the evaluated context and execution budgets. 
{\method} improves FRR while maintaining comparable Micro Precision to the strongest baselines and using fewer execution tokens than AST+GraphRAG, suggesting that the gain is not solely explained by exposing more repository content.
{\method} uses fewer execution tokens than AST+GraphRAG while achieving higher FRR, further undermining the explanation that the gain simply comes from exposing more repository content.
FRR should nonetheless be interpreted jointly with precision and cost.


\begin{tcolorbox}[size=title]
\textit{\textbf{Answer for RQ1:} On the medium-scale benchmark, {\method} obtains the highest FRR among the compared methods without relying on persistent offline indexes or pre-built repository graphs. Under the controlled setting, it provides a favorable recall–cost trade-off compared with the evaluated RAG-style alternatives.}
\end{tcolorbox}

\noindent\textbf{(2) RQ2: Repository Understanding in AI Coding Systems.}
\begin{figure*}[t]
\centering

\subfloat[Large Projects, Run 1]{
    \label{fig:large_run1}
    \includegraphics[width=0.27\textwidth]{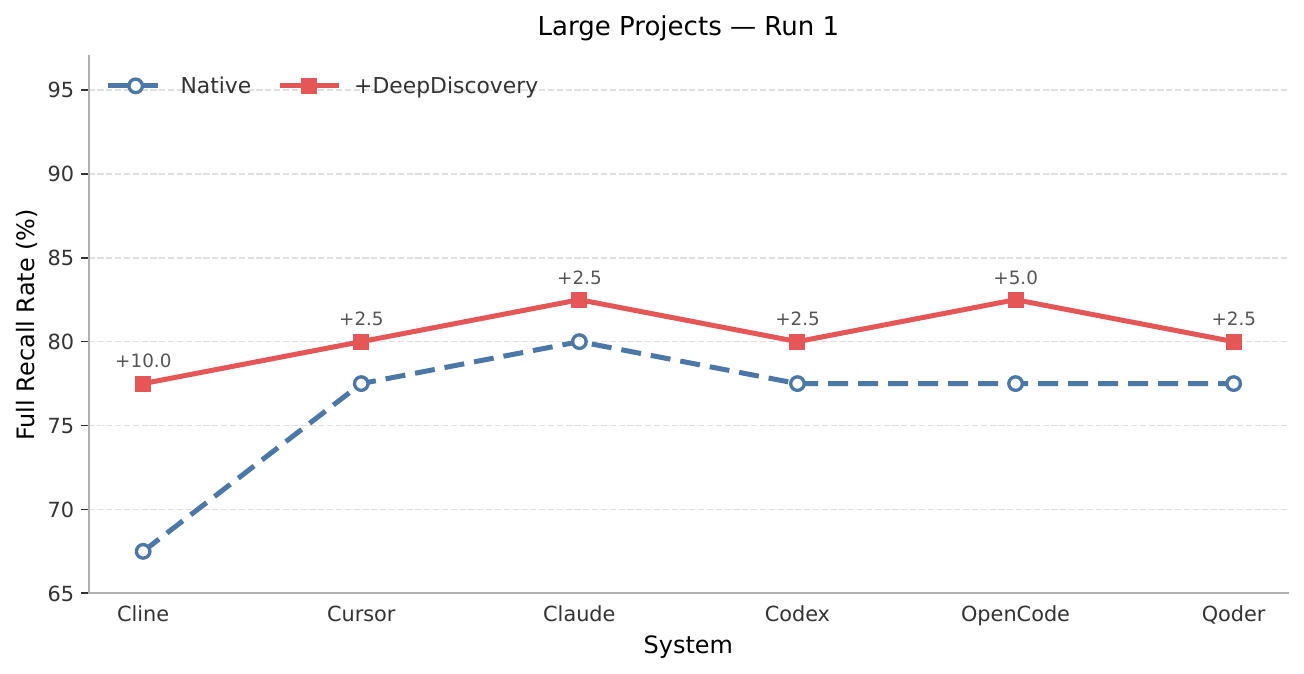}
}
\hfill
\subfloat[Large Projects, Run 2]{
    \label{fig:large_run2}
    \includegraphics[width=0.27\textwidth]{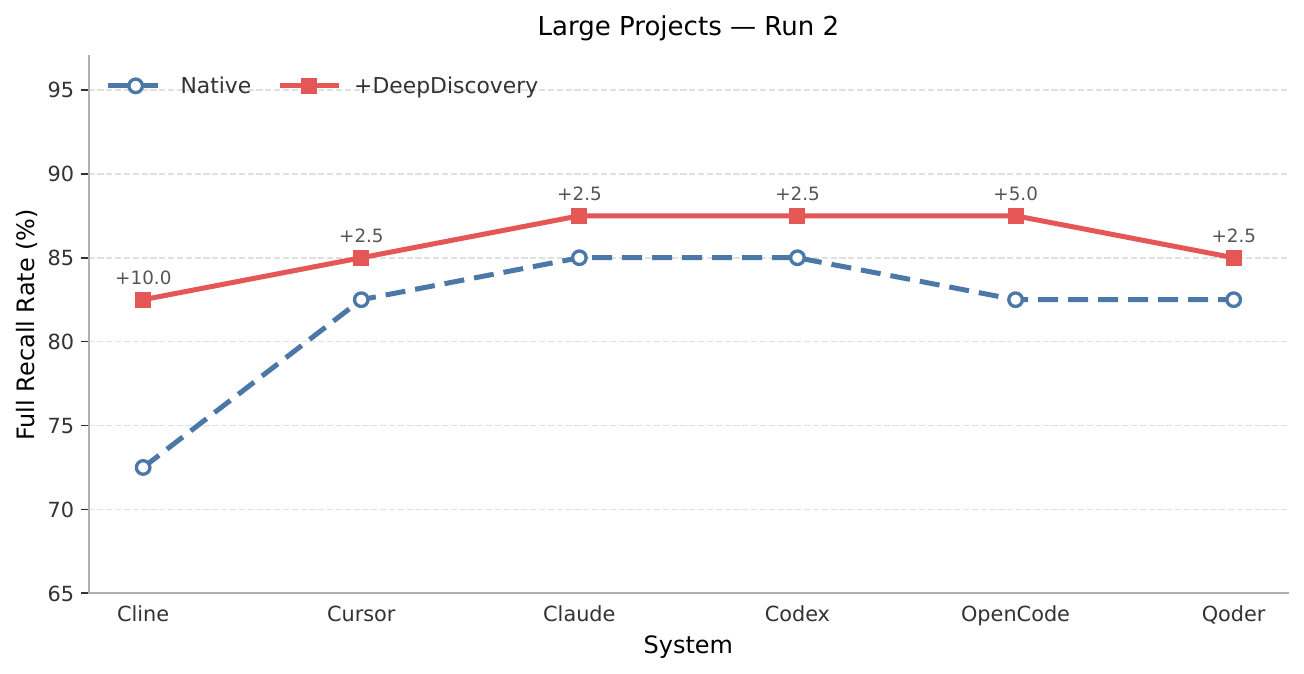}
}
\hfill
\subfloat[Large Projects, Run 3]{
    \label{fig:large_run3}
    \includegraphics[width=0.27\textwidth]{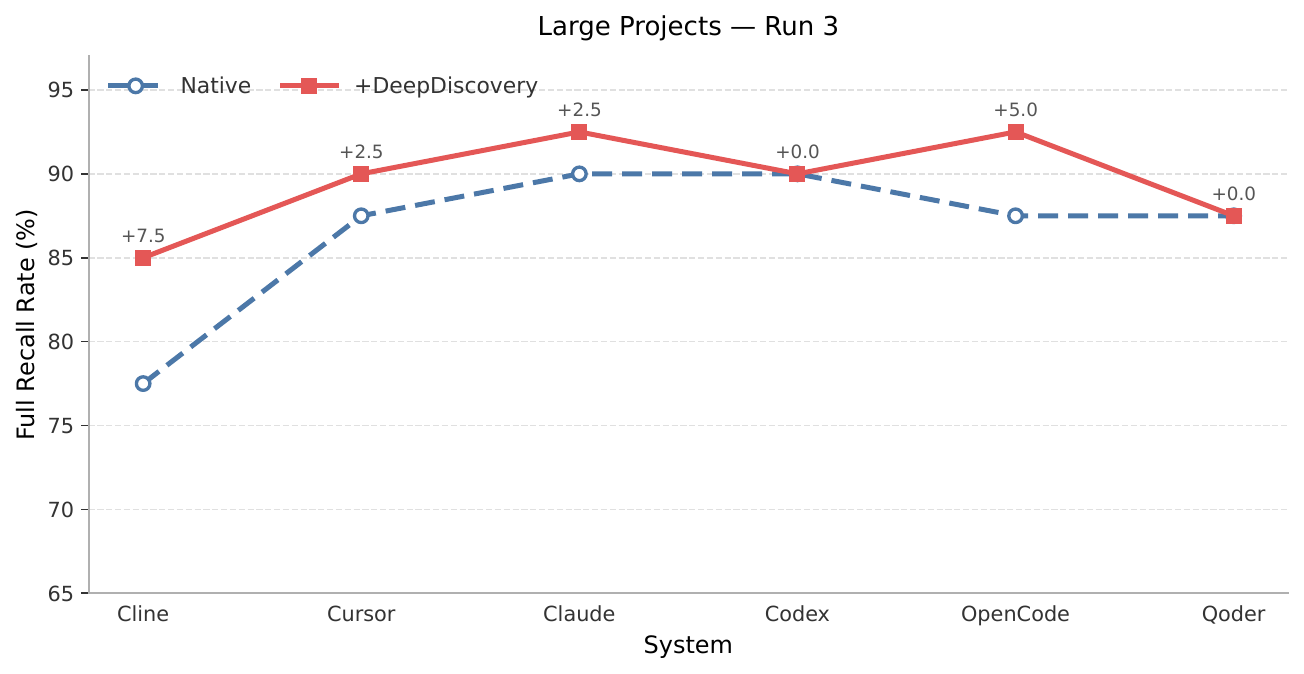}
}

\vspace{0.2em}

\subfloat[Medium-scale Projects, Run 1]{
    \label{fig:medium_run1}
    \includegraphics[width=0.27\textwidth]{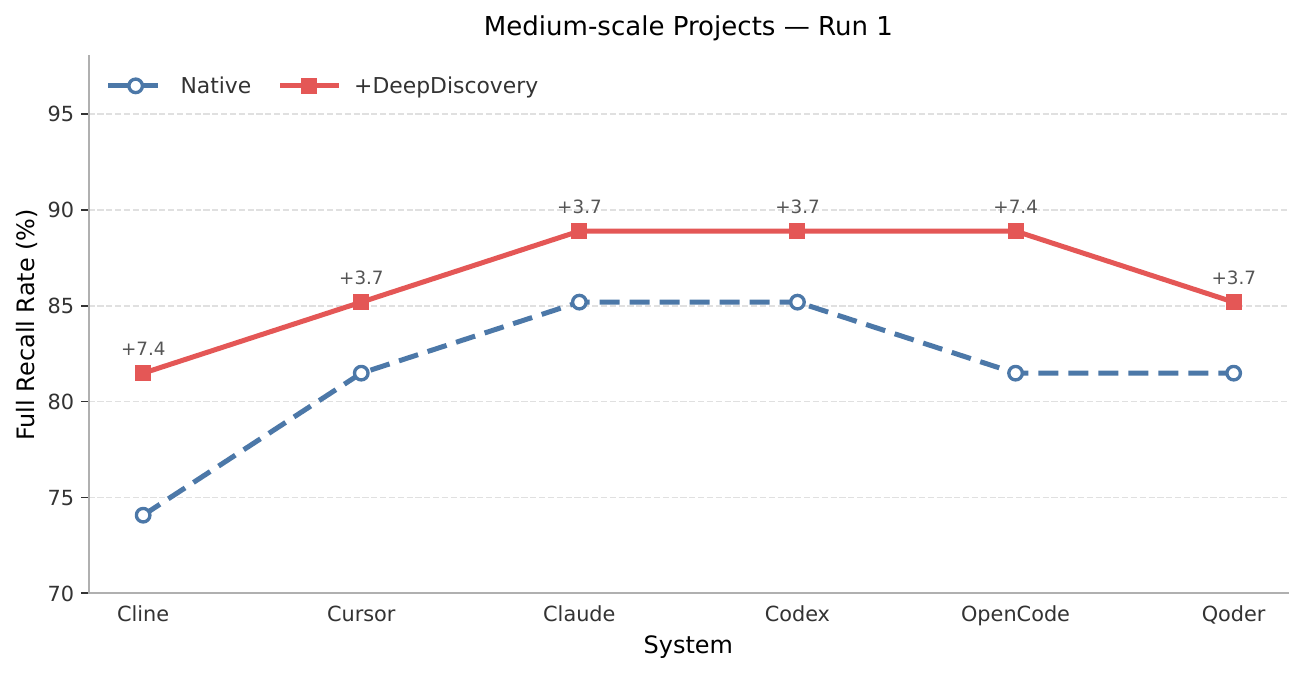}
}
\hfill
\subfloat[Medium-scale Projects, Run 2]{
    \label{fig:medium_run2}
    \includegraphics[width=0.27\textwidth]{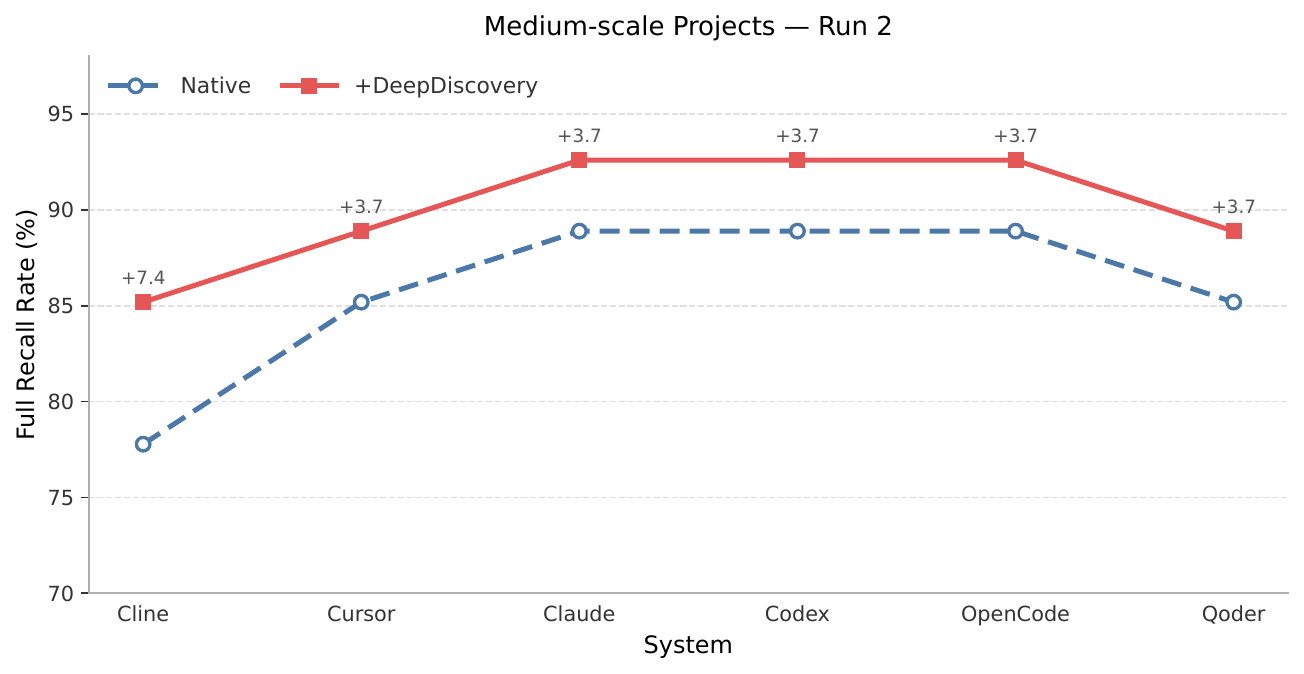}
}
\hfill
\subfloat[Medium-scale Projects, Run 3]{
    \label{fig:medium_run3}
    \includegraphics[width=0.27\textwidth]{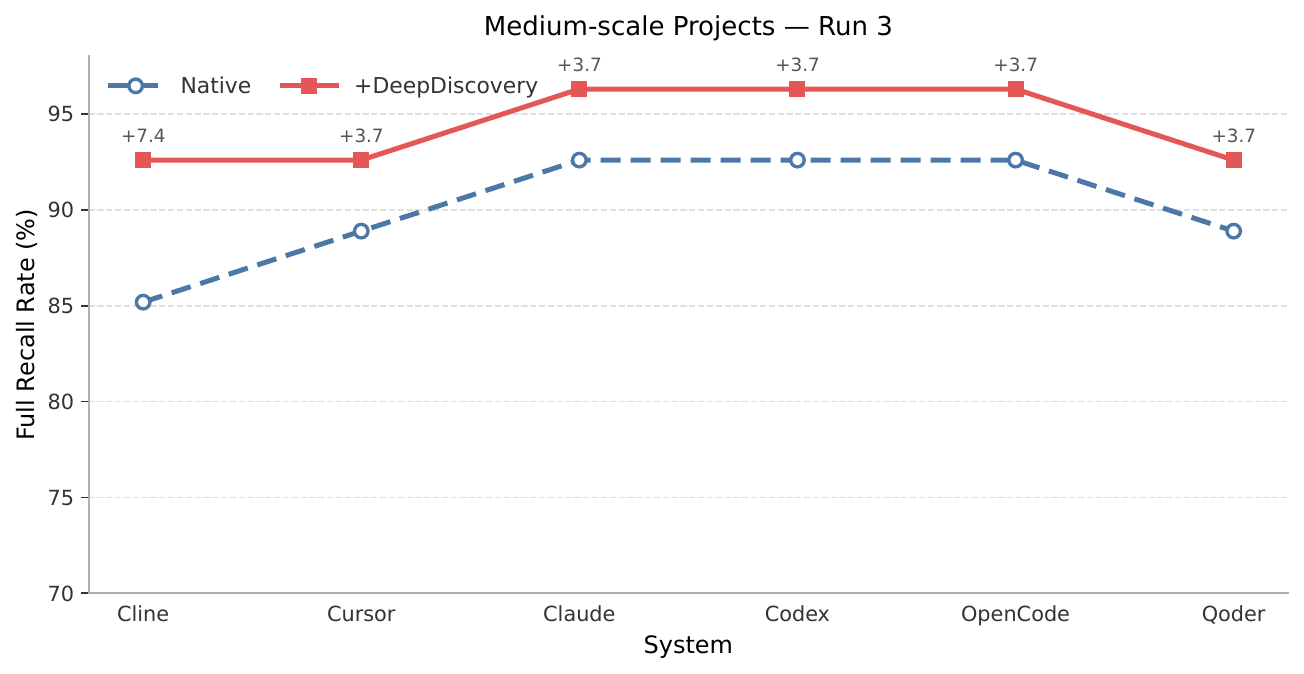}
}

\caption{Run-level Full Recall Rate across project scales and repeated executions. The first row reports large-subproject results, and the second row reports medium-scale results; each column corresponds to one repeated run. In nearly all systems and runs, the {\method}-enhanced setting achieves higher FRR than the native setting, indicating consistent gains across scales and repetitions rather than run-specific effects.}
\label{fig:frr_runs_all}
\end{figure*}

Across all evaluated systems, the most consistent improvement is observed in FRR. 
Table~\ref{tab:main_compare} summarizes the aggregate results and shows run stability and metric-wise gains.
On large subprojects, FRR improves for all six systems, with the largest gains for Cline and OpenCode. 
The same pattern holds on medium-scale subprojects, with smaller margins.
These gains reflect improved recovery of complete task-relevant file sets. 
Figure~\ref{fig:frr_runs_all} shows that the improvement is stable across runs and generally larger on large subprojects.

This pattern is consistent with the interpretation that systems with weaker native repository search and context construction capabilities have more headroom to benefit from stronger task-level context recovery.
Table~\ref{tab:main_compare} further shows that gains are concentrated more on MR than on MP, indicating that {\method} primarily improves completeness rather than aggressive pruning.
Overall, {\method} improves repository-understanding quality across real AI coding systems, with the clearest gains on FRR.


\begin{tcolorbox}[size=title]
\textit{\textbf{Answer for RQ2:} In the evaluated AI coding systems, {\method} improves FRR in all reported settings and generally improves Micro Recall. The evidence is strongest for completeness-oriented recovery of task-relevant files, while precision changes are smaller and occasionally mixed.}
\end{tcolorbox}

\noindent\textbf{(3) RQ3: Ablation and Component Analysis}

Because the Inference stage relies on anchors produced by Location, removing the entire Location stage would not yield a meaningful executable variant. We therefore perform targeted ablations that preserve the semantics of the pipeline. Specifically, we compare the full system with four reduced variants: \textbf{Location-only}, \textbf{w/o Adaptive Compression}, \textbf{w/o Implicit Relations}, and \textbf{w/o Metadata-First}.

Table~\ref{tab:ablation_main} shows that all reduced variants underperform the full system, indicating that each major component contributes materially to task-level context recovery. \textbf{Location-only} lowers FRR from 85.8\% to 80.0\%, showing that anchor localization alone is insufficient for recovering complete implementation paths without subsequent expansion. Removing adaptive compression reduces FRR to 81.5\% and MR to 83.9\%, suggesting that structure-preserving compression helps retain repository signals at a granularity useful for both anchor identification and downstream expansion. Removing implicit relations reduces FRR to 83.3\% and MR to 87.4\%, confirming that many task-relevant files are connected through framework-mediated or configuration-level links beyond explicit dependencies. Removing metadata-first context construction reduces FRR to 82.2\% and MR to 85.7\%, suggesting that preserving structural coverage before selective full-text promotion is more effective than exposing detailed content too early.

Overall, the gains are concentrated more on FRR and MR than on MP, which is consistent with the recall-oriented design of {\method}. 

\begin{tcolorbox}[size=title]
\textit{\textbf{Answer for RQ3:} All major components of {\method} contribute to task-level context recovery. Removing adaptive compression, implicit relations, metadata-first context construction, or inference-based expansion consistently reduces recovery quality, and the full pipeline achieves the best overall performance.}
\end{tcolorbox}

\begin{table}[t]
    \caption{Ablation results on the large-subproject benchmark under a fixed host-system setting. The full \textit{Location--Inference} pipeline outperforms all reduced variants, indicating that each major component contributes non-trivially to task-level context recovery.}
    \label{tab:ablation_main}
    \centering
    \footnotesize
    \setlength{\tabcolsep}{5pt}
    \begin{tabular}{lccc}
        \toprule
        Variant & FRR & MR & MP \\
        \midrule
        Location-only            & 80.0\% & 85.8\% & 9.5\% \\
        w/o Adaptive Compression & 81.5\% & 83.9\% & 9.0\% \\
        w/o Implicit Relations   & 83.3\% & 87.4\% & 9.4\% \\
        w/o Metadata-First       & 82.2\% & 85.7\% & 9.2\% \\
        {\method} (full)         & \textbf{85.8\%} & \textbf{89.6\%} & \textbf{9.3\%} \\
    \bottomrule
    \end{tabular}
\end{table}

\noindent\textbf{(4) RQ4: End-to-End Task Solving Capability.}

\begin{figure}[t]
    \centering
    \includegraphics[width=0.85\linewidth]{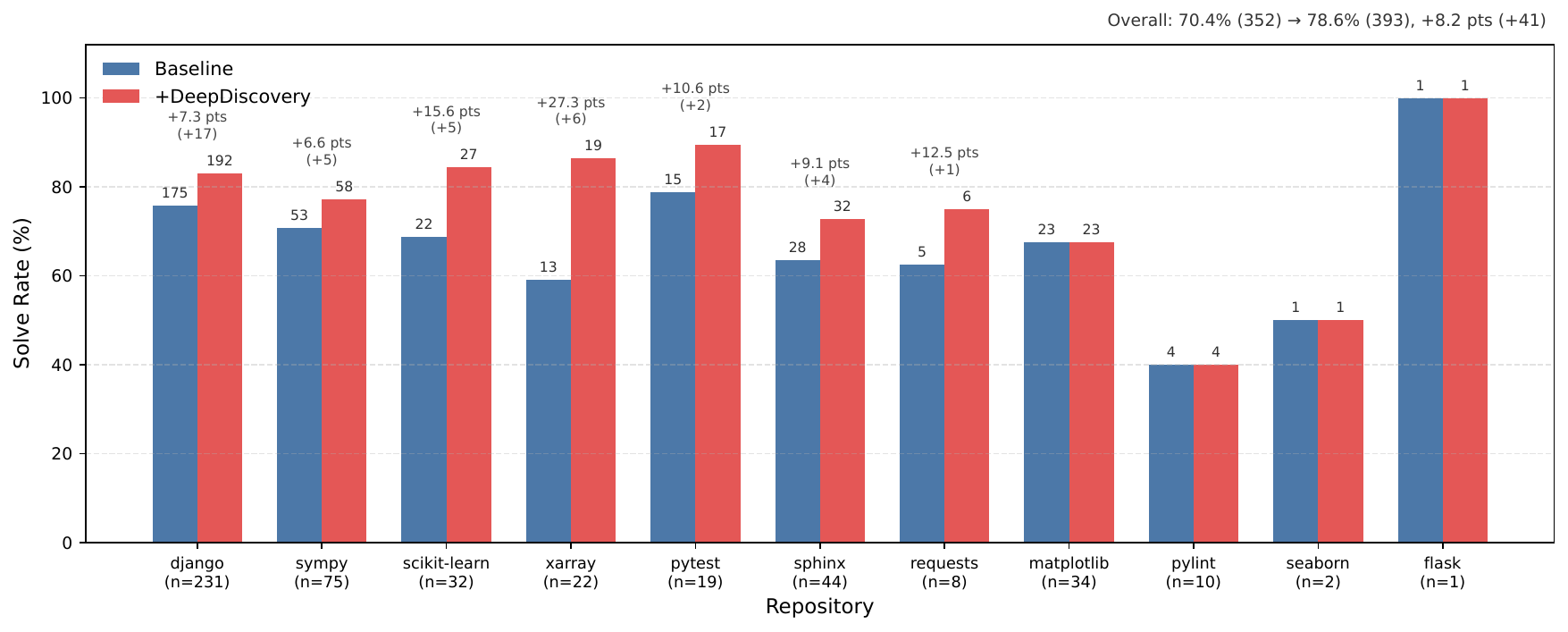}
    \caption{Solve Rate on SWE-bench Verified by repository. Bar heights show Solve Rate, numbers above bars show resolved cases, and gain annotations show percentage-point improvements and added resolved cases. {\method} improves end-to-end performance across several major repositories.}
    \label{fig:swebench_repo_solve}
\end{figure}
Figure~\ref{fig:swebench_repo_solve} shows the repository-wise results.
Overall, the {\method}-enhanced system solves 393 of 500 tasks (78.6\%), compared with 352 of 500 (70.4\%) for the baseline, for an absolute gain of 8.2 percentage points (+41 tasks).
The gain is statistically significant under a paired McNemar test on per-instance outcomes ($p<0.01$).
This matched comparison evaluates the effect of replacing the task-time repository-context construction component with the Location–Inference context package while keeping the model, prompting strategy, tool interface, execution policy, and evaluation environment fixed. 
Concretely, the baseline system uses its native repository search and context assembly procedure, whereas the {\method}-enhanced system receives the structured context package produced by {\method} in place of that native context-construction output.

The repository-wise breakdown shows that gains are concentrated in projects whose tasks more often require cross-file, cross-module, or configuration-sensitive reasoning.
In particular, {\method} solves 17 additional tasks on django, 6 on xarray, 5 on sympy, 5 on scikit-learn, 4 on sphinx, and 2 on pytest, with an additional gain on requests.
By contrast, gains are limited or absent on more localized subsets, consistent with the zero-gain cases on matplotlib, pylint, seaborn, and flask.
These results align with the central objective of DEEPDISCOVERY: improving downstream task solving by recovering structured task-level context when relevant evidence is distributed across the repository.

\begin{tcolorbox}[size=title]
\textit{\textbf{Answer for RQ4:} {\method} improves end-to-end task solving on realistic software engineering tasks. Under the matched comparison, it increases Solve Rate significantly, especially on tasks requiring broader repository-level reasoning.}
\end{tcolorbox}

\noindent\textbf{(5) RQ5: Robustness Across Scales, Models, and Systems.}

\begin{table}[t]
\caption{Cross-system robustness of {\method} in terms of Full Recall Rate gain across project scales. All evaluated systems obtain positive gains on both large and medium-scale subprojects.}
\label{tab:rq4_cross_system}
\centering
\footnotesize
\setlength{\tabcolsep}{4pt}
\begin{tabular}{lccc}
\toprule
System & Large Gain & Medium Gain & Consistency \\
\midrule
Cline        & +9.2  & +7.4 & 2/2 positive \\
Cursor       & +2.5  & +3.7 & 2/2 positive \\
Claude Code  & +2.5  & +3.7 & 2/2 positive \\
Codex        & +1.6  & +3.7 & 2/2 positive \\
OpenCode     & +5.0  & +4.9 & 2/2 positive \\
Qoder        & +1.7  & +3.7 & 2/2 positive \\
\bottomrule
\end{tabular}
\end{table}

Finally, we examine whether the gains of {\method} remain consistent across heterogeneous AI coding systems, project scales, and model backends. The evaluated systems differ in both underlying models and agent workflows. 
As shown in Table~\ref{tab:rq4_cross_system}, all six systems achieve positive FRR gains on both large and medium-scale subprojects. 
The gains are largest for Cline and OpenCode, while Claude Code and Codex still show smaller but consistent improvements. 
These results suggest that the benefit of {\method} is robust across diverse native backends rather than tied to a particular agent workflow. 
Given that these systems already employ strong but different repository search and context-construction strategies, the observed gains are more likely to come from improved task-level repository understanding than from weak baselines. 
We also observe that stronger foundation models tend to achieve higher absolute FRR under the same integration setting, suggesting that repository understanding and model reasoning are complementary.

\begin{tcolorbox}[size=title]
\textit{\textbf{Answer for RQ5:} The observed FRR gains are positive across the evaluated systems, project scales, and model backends, suggesting that the Location–Inference workflow is not tied to a single host agent in the studied settings.}
\end{tcolorbox}

\section{Discussion}
\vspace{-0.2em}
\subsection{Efficiency and Practical Considerations}

We next consider whether the quality gains of {\method} come at an acceptable practical cost. 
{\method} is designed for settings where repository freshness and context budget matter. Because it operates on the current repository snapshot, it avoids maintaining persistent repository-wide indexes or pre-built graphs. This is advantageous for fast-changing repositories, although offline artifacts may still be preferable when repositories are stable and many repeated queries amortize preprocessing cost.

The staged workflow also improves the latency–quality trade-off. In our measurements, coarse localization through repository compression takes about 70 seconds on average, compared with about 225 seconds for broad local search alone. {\method} therefore uses compression to narrow the search space, then applies relation-guided expansion and selective full-text promotion only where additional evidence is needed. This design adds control logic and metadata-management overhead, but the measured overhead is acceptable in the evaluated settings.
\subsection{Error modes and scope.}

Although {\method} improves task-level file recovery in most evaluated settings, it is not uniformly beneficial on every task. 
Our observations suggest three main categories of failure modes.
First, anchor localization can still be misled by weak, underspecified, or ambiguous task descriptions, especially in repositories with inconsistent naming conventions or weak organizational signals. 
In such cases, the system may begin expansion from suboptimal entry points, which can limit the quality of the recovered implementation path.
Second, implicit-relation expansion can introduce structurally adjacent but non-essential files. 
This behavior is partly consistent with the recall-oriented design of {\method}: improving implementation-path completeness may also bring in a limited amount of nearby non-gold context. 
While this trade-off is often acceptable in workflows where missing a critical bridge or configuration file is especially costly, it can reduce precision in settings with very tight context budgets.
Third, metadata-first exploration can under-expose fine-grained implementation details when the promotion decision is too conservative. 
In such cases, metadata may preserve structural coverage but still fail to surface the detailed code evidence needed for downstream reasoning or modification.
At present, {\method} is best viewed as a recall-oriented repository-understanding enhancement.

\vspace{-0.2em}
\section{Conclusion}
\label{sec:conclusion}
We presented {\method}, a Location–Inference framework that treats repository understanding as budgeted recovery of structured task context. 
By combining anchor localization, multi-relational expansion, and metadata-first context construction, {\method} improves recovery of task-relevant files and relations across controlled repository-understanding evaluations, industrial-scale coding-system integrations, and SWE-bench Verified. 
These results indicate that coding agents benefit from repository-understanding components that recover connected implementation paths rather than isolated relevant fragments. 
Future work will study broader cross-organization settings and finer-grained policies for context promotion and budget allocation.

\bibliographystyle{IEEEtran}
\bibliography{ref}

\end{document}